\documentclass[twocolumn]{article}
\usepackage{imr}
\usepackage{graphicx}
\usepackage{lipsum}  

\usepackage{algorithm}
\usepackage{algpseudocode}

\usepackage{amsmath}
\usepackage{mathtools}  % brings in amsmath, also some improvements
\usepackage{amssymb} % brings in amsfonts, incl \square

\usepackage{hyperref}
\usepackage{caption}

\algdef{SE}[DOWHILE]{Do}{doWhile}{\algorithmicdo}[1]{\algorithmicwhile\ #1}%

\algblockdefx[kernel]{BKernel}{EKernel}{\textbf{kernel }}{\textbf{end kernel}}
\algdef{SE}[DOWHILE]{Do}{doWhile}{\algorithmicdo}[1]{\algorithmicwhile\ #1}%

\begin{document}

\title{\uppercase{GPU Parallel algorithm for the generation of polygonal meshes based on terminal-edge regions}}
%\author{Double Blind peer review}
\title{\uppercase{GPU Parallel algorithm for the generation of polygonal meshes based on terminal-edge regions}}
\author{Sergio Salinas$^1$ \and José Ojeda$^2$ \and Nancy Hitschfeld $^3$ \and Alejandro Ortiz-Bernardin$^4$ }
\date{
$^1$University of Chile, Santiago, RM, Chile. ssalinas@dcc.uchile.cl\\
$^2$University of Chile, Santiago, RM, Chile. Jojeda@dcc.uchile.cl\\
$^3$University of Chile, Santiago, RM, Chile. nancy@dcc.uchile.cl\\
$^4$University of Chile, Santiago, RM, Chile. aortizb@ing.uchile.cl\\
}

%% \abstract{*} and \keywords{*} must be before \maketitle.
\abstract{This paper\footnote{Workshop submitted to the  SIAM International Meshing Roundtable Workshop 2022 (SIAM IMR22), Feb. 22-25, 2022, online-only conference. This workshop describes an ongoing work that when it is finished it will be sent to a journal or
conference.} presents a  GPU parallel algorithm  to generate a new kind of polygonal meshes obtained from Delaunay triangulations. To generate the polygonal mesh, the algorithm first uses a classification system to label each edge of an input triangulation; second it builds polygons (simple or not) from terminal-edge regions using the label system, and third it transforms each non-simple polygon from the previous phase into simple ones, convex or not convex polygons. We show some preliminary experiments to test the scalability of the algorithm and   compare it   with the sequential version. We also run a very simple test to show that these meshes can be useful for the virtual element method.}% Both types are permitted. We analyse which kind of non simple polygons appear and show the algorithm to divide them into simple ones.}


\maketitle
\thispagestyle{empty}
\pagestyle{empty}

\section{Introduction}

%es indirect method
%comparar secuencial
%comparar optimización por trivertexe

%\section{Related work}

%Generación de mallas

Polygonal mesh generation is an area of study broadly researched due to their applications in computer graphics \cite{PolygonalmeshGrapfics}, geographic information systems \cite{PrinciplesGIS} and finite element methods (FEM) \cite{HOLE198827}, among others. As a result of the FEM research, several requirements on the shape of the basic cells have been established  as quality shape criteria. Typical meshes tend to contain only triangles or quadrilaterals; the big exception is the use of Voronoi regions (convex polygons) as basic cells~\cite{GHOSH199433}. %Few researchers \cite{} have looked for other kind of convex polygons. 
However, in the recent years, the virtual element method (VEM) \cite{Beir2013BasicPO} has showed to work with polygons convex and non convex \cite{CHI2017148, PARK2019669}, so a new area of study to generate  quality meshes for the VEM has begun \cite{ATTENE20211392, Vemandthemesh}.

Common methods to generate  unstructured mesh generation are the Delaunay methods \cite{cheng2013delaunay}, Voronoi diagram methods \cite{YAN2013843, 2dcentroidalvoro, talischi2012polymesher}, Advancing front method \cite{lohner1996progress}, quadtree based methods \cite{bommes2013quad} and hybrids methods \cite{owen1999q}. In general, meshing algorithms can be classified into two groups \cite{owen1998survey, johnen2016indirect}: (i) direct algorithms:  meshes are  generated from the input geometry, and (ii) indirect algorithms: meshes are generated starting from an input mesh, tipically an initial triangle mesh. Indirect methods is a common approach  to generate quadrilateral meshes by mixing triangles of an initial triangulation ~\cite{LeetreetoQuad, BlossonQuad, Merhof2007Aniso-5662}. The advantage of using indirect methods is that triangular meshes are pretty simple to generate because there are several robust and well-studied programs for generating triangulations~\cite{qhull, triangle2d, Detri2}.

%Generación de mallas en cuda
There are a widely research in parallel mesh generation \cite{ParallelMeshGeneration, Paralleladvancingfront, ParallelDelaunay, Bernparallelquadtree, barragan2000two}. But research in 2D mesh generation taking advantage of GPU parallelization is not so common. There are well known algorithms to generate Voronoi diagrams \cite{VoronoiGPU, VoroGPU2, VoroGPU}, Delaunay triangulations \cite{Delaunaygpu, 6361389} and quadtree based meshes \cite{kelly2011quad} in GPU using direct methods. In the case of indirect methods an interesting algorithm to generate a Delaunay triangulation via edge-flipping was proposed in \cite{NavarroHS13}.

\begin{figure}
    \centering
  %\includesvg[width=1.\linewidth, angle=180]{pygal_cuyi2_81.svg}
  \includegraphics[width=1.02\linewidth]{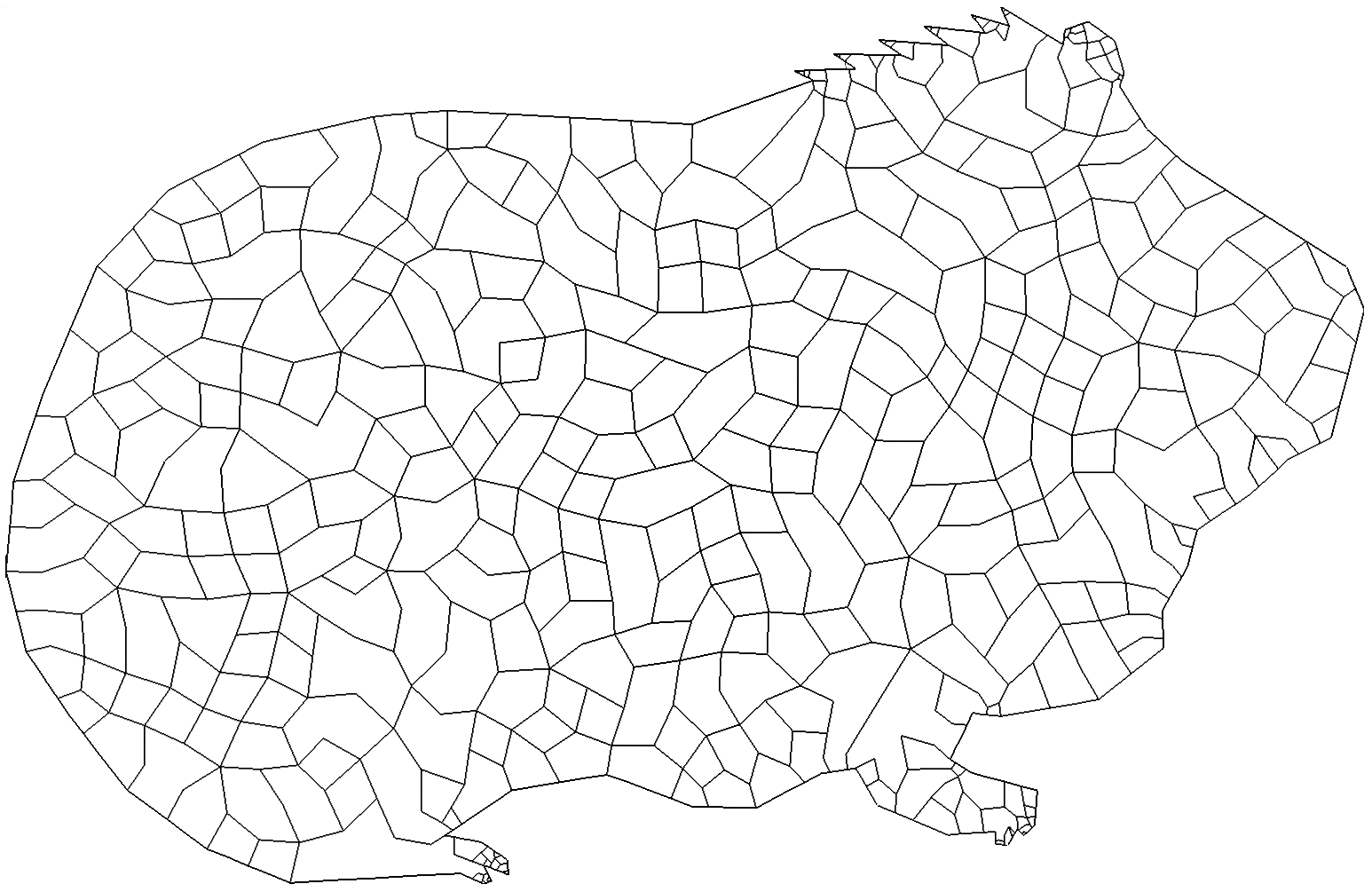}
  \caption{Polygonal mesh of a guinea pig using terminal-edge regions as polygons. The initial triangulation of the PLSG was made using Pygalmesh \cite{pygalmesh} with the constrain of max edge size of $100$. The resulting input triangulation contains $1330$ triangles, $819$ vertices and $2148$ edges. The output polygonal mesh contains $363$ polygons, $819$ vertices and $1441$ edges.   }
  \label{fig:PLSGEXample}
\end{figure}

In this paper we present a GPU parallel algorithm  to generate a new kind of 2D polygonal meshes using the concept of terminal-edge regions.  This algorithm belongs to the indirect methods. An example of a polygonal mesh generated by the algorithm is shown in Figure \ref{fig:PLSGEXample}.  The  data structure and several kernels to work with the input triangulation and build the  arbitrary shape polygons in GPU are described. In particular,  the  AtomicAdd function is used to  manage the concurrent saving of polygons in an array. 

This paper is organized as follows: Section \ref{sec:background} introduces the concepts necessary to understand the algorithm. Section \ref{sec:datastruct} describes the data structure uses to represent the input triangulation and the polygonal mesh in the GPU. Section \ref{sec:algorithm} presents the sequential algorithm and the parallel algorithm. Section \ref{sec:Experiments} shows performance experiments of the sequential and parallel algorithms. Finally, Section \ref{sec:conclusions} presents the conclusions and  the ongoing work in the parallel algorithm.

\section{Data structure}
\label{sec:datastruct}

The GPU algorithm receives as input a triangulation. Since GPUs (and their programming models) are designed to work efficiently with arrays, we decided to represent the triangulation with an indexed data structure with neighbor indexation~\cite{DataStructSurvey}. This representation uses 4 arrays, a {\sc Vertex array}, a {\sc Triangle array}, a {\sc Neighbor array} and a {\sc Trivertex array}.

\begin{itemize}
    \item The {\sc Vertex array} stores each point of the triangulation in pairs, each two elements \texttt{V[2$\cdot i + 0$}], \texttt{V[2$\cdot i + 1$}] in the array is the coordinate to the i-th point in the triangulation.
    \item The {\sc Triangle array} contains indices to the {\sc Vertex array}, each $3$ elements \texttt{T[3$\cdot i + 0$}], \texttt{T[3$\cdot i + 1$}], \texttt{T[3$\cdot i + 2$}] are indices to the vertices of the i-th triangle in the triangulation. 
    \item The {\sc Neighbor array} stores indices to the {\sc Triangle array}, each three indices \texttt{N[3$\cdot i + 0$}], \texttt{N[3$\cdot i + 1$}], \texttt{N[3$\cdot i + 2$}] are the indices to a triangle neighbor to the triangle $i$, the values 0, 1 and 2 represents the edges of $i$.
    \item The {\sc Trivertex array}, stores the index of one triangle incident to each vertex $i$ in the  {\sc Vertex array}.  This array is optional and  is only used to accelerate the Reparation phase in section \ref{subsubsec:reparation}. 
    
\end{itemize}

This representation has the advantage that it can be easily allocated in the CUDA's global device memory, it allows us to assign  a unique triangle to each thread and it facilitates the traversal between neighbor triangles in constant time. Also, several programs as QHull~\cite{qhull}, Triangle~\cite{triangle2d} and Detri2~\cite{Detri2} store this representation as output, except for the Trivertex array that can be obtained from the Vertex array and the Triangle array, therefore we can integrate external libraries to generate the initial triangulation.

Polygons are represented as a set of vertices, so to store them, the algorithm uses an array of arbitrary length of indices to the {\sc Vertex array}.

The output polygonal $\tau' = (V,E)$ mesh generated by the algorithm is represented as two arrays in the global memory device. The first one is the  \texttt{Mesh array}, this array represents each polygon in $\tau'$ as a number that indicates its length and a set of vertices in counter-clock wise that are part of the polygon. The second one is the \texttt{Position array}, that stores the index to the position of the first element of each polygon in the \texttt{Mesh array}.

%set of polygons, this is store as a \texttt{Mesh array} in the global memory device, where to store each polygon $P$, the first element store, in the mesh, is the length of $P$ and after that the set of vertices $P_i$ that represent $P$.

Other two important data structures  are: (i) \texttt{Seed array}: a bit array whose values are true only associated to the indices of the algorithm to generate new polygons, (ii) \texttt{Max array}: given a triangle $i$  in the {\sc Neighbor array}, \texttt{Max array} stores the index $j$ if the implicit edge stored at $3i+j$ contains the longest-edge of the triangle $i$. Both arrays are stored in the global device memory.

\section{Algorithm}
\label{sec:algorithm}

\subsection{Sequential algorithm}
\label{subsec:seq}

The sequential algorithm is actually submitted to a journal for publication; an early pre-print version can be seen in \cite{salinasstudying}.

\subsection{GPU parallel algorithm}
\label{subsec:GPU}

The general algorithm is shown in \ref{algo:generalalgorithm}. The algorithm takes as input a triangulation $\tau_L(V,E)$ and generates as output a polygonal mesh. To do so, the algorithm consists of three main phases: label each edge and triangle of $\tau$, traversal inside terminal-edge regions and repair non-simple polygons generated from the last phase. The first two phases have the advantage that they are extremely  data-parallel; it is possible to work with each element in $\tau$ independently, so these phases take advantage of the data parallelism capabilities of the GPU programming model. The Reparation phase needs to split a non-simple polygon into two until all their barrier-edge tips are removed, i.e., transformed in  frontier-edge endpoints. Since a non-simple polygon can have more than one barrier-edge, this phase will be repeated until all of them are absorbed by a frontier-edge.  The experiments described in \ref{sec:Experiments} show us that there are few non-simple polygons after the Traversal phase. Despite the  parallel algorithm we designed for  this phase,  it seems better to use just the sequential version of this phase.

\begin{algorithm}
    \caption{Polygonal Mesh generator}\label{algo:generalalgorithm}
    \begin{algorithmic}[1]
    \Require Initial Triangulation  $\tau = (V,E)$
    \Ensure Polygonal Mesh $M$
    
    \State Label triangulation $\tau$ \Comment{Algorithm \ref{algo:labelphase}}
    \State Generate polygons  \Comment{Algorithm \ref{algo:TravelPhase}}
    \Do  
        \State Remove barrier-edge tips \Comment{Algorithm \ref{algo:ReparationPhase}}     
    \doWhile{$M$ contains Barrier-edge tips}
    \State \Return Polygonal Mesh
    \end{algorithmic}
\end{algorithm}

\subsubsection{Label Phase}

This phase reads a triangulation $\tau = (V,E)$ as input. The objective  is to label frontier-edges in $\tau$ to identify the boundary of each terminal-edge region $R_i$ in $\tau$ and to select one triangle of each $R_i$ to be used as generator of new polygons in the next phase, the Traversal phase. This phase calls three kernels: \texttt{LabelMax}, \texttt{LabelSeed} and \texttt{LabelFrontier}. Those kernels are shown in Algorithm \ref{algo:labelphase} and they are called consecutively.

\begin{algorithm}
    \caption{Label phase's kernels}\label{algo:labelphase}
    \begin{algorithmic}[0]
    \Require Initial Triangulation  $\tau = (V,E)$
    \Ensure Labeled  triangulation $\tau_L = (V,E)$
    \BKernel{\texttt{LabelMax}($\tau = (V,E)$)}
    \ForAll{triangle $t_i$ in $\tau$ \textbf{in parallel} }\label{algolabel:TriangleIteration}
        \State Label the longest-edge of $t_i$
    \EndFor
    \EKernel
    \BKernel{\texttt{LabelSeed}($\tau = (V,E)$)}
    \ForAll{edge $e_i \in \tau$ \textbf{in parallel}} \label{algolabel:Edgeiteration}
        \If{$e_i$ is border terminal-edge}
            \State Store adjacent triangle as seed triangle
        \EndIf
        \State Be $t_a$ and $t_b$ triangles that share $e_i$
        \If{$e_i$ is terminal-edge \textbf{and} $a < b$}
                \State Store $t_a$ as seed triangle
        \EndIf
    \EndFor
    \EKernel
    \BKernel{\texttt{LabelFrontier}($\tau = (V,E))$}
    \ForAll{edge $e_i \in \tau$ \textbf{in parallel}} \label{algolabel:Edgeiteration2}
        \State Be $t_a$ and $t_b$ triangles that share $e_i$
        \If{$e_i$ neither the longest-edge of $t_a$ nor $t_b$ \textbf{or} $e_i$ is border-edge} \label{algolabel:label}
            \State Label $e_i$ as frontier-edge
        \EndIf    
    \EndFor
    \EKernel
    \end{algorithmic}
\end{algorithm}

The first kernel, \texttt{LabelMax} assigns a triangle $t$ of $\tau$ to each thread,  then it calculates its longest-edge  and stores its index ($0$, $1$ or $2$) in the \texttt{Max Array}.

The second kernel, \ref{algolabel:Edgeiteration}, \texttt{LabelSeed}, assigns an edge $e$ of $\tau$ to each thread.  It compares two triangles $t_a$ and $t_b$ that share $e$ and checks if in both triangles, $e$ is the longest-edge. In that case, $e$ is a terminal-edge and the triangle with lower index is set to \texttt{True} in \texttt{Seed array}, for the generation of a new polygon in the next phase. In case $e$ is both a boundary edge and the longest-edge of the triangle $t$, then $e$ is a boundary terminal-edge and $t$ is stored as seed triangle in the \texttt{Seed array}.

\begin{figure}
  \includegraphics[width=\linewidth]{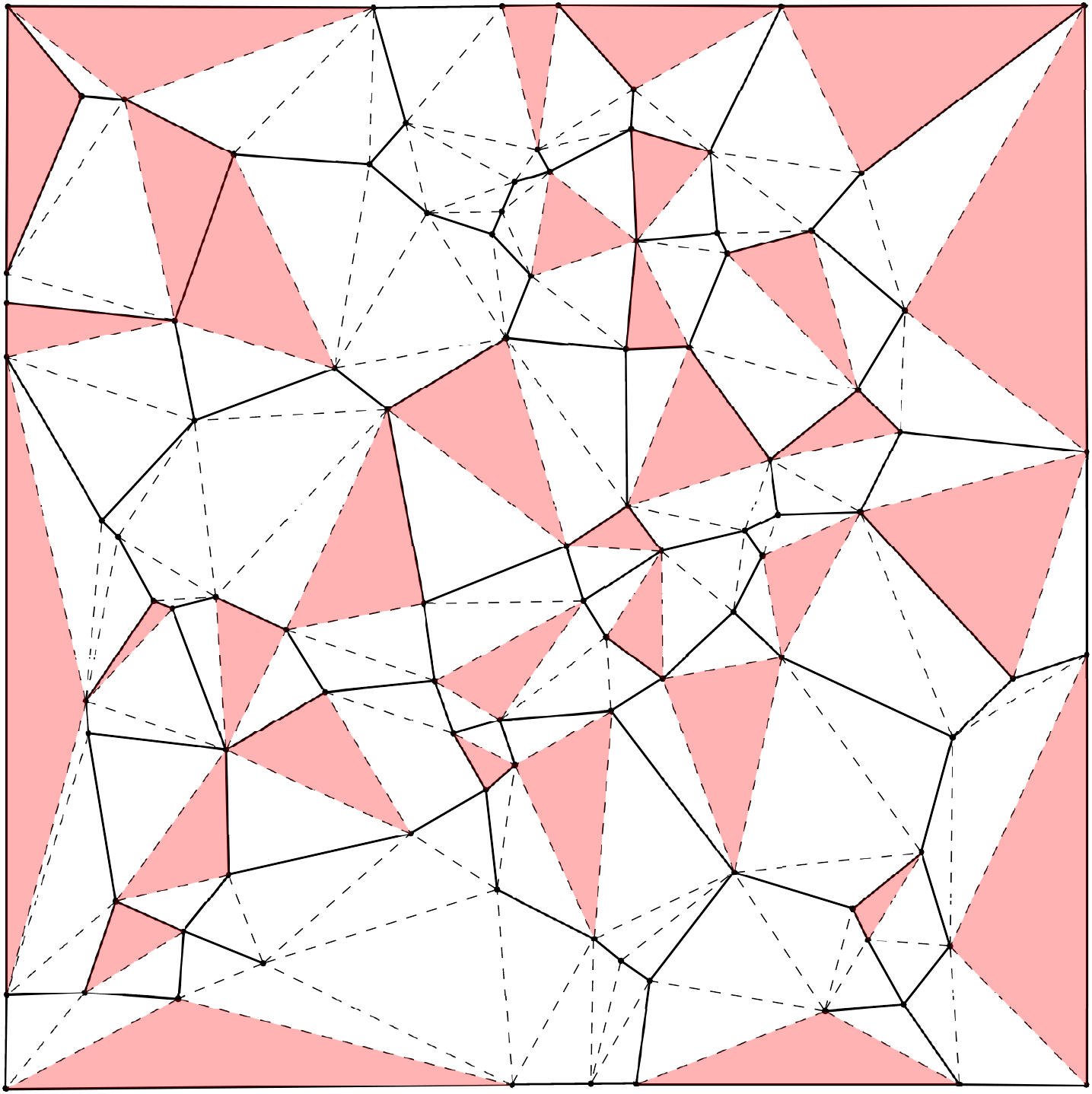}
  \caption{Labeled triangulation generated by the Algorithm \ref{algo:labelphase}. Solid lines are frontier-edge, dashed lined are internal-edges and terminal-edges. Red triangles are seed triangles.}
  \label{fig:labeledtriangluation}
\end{figure}

The third kernel, \ref{algolabel:Edgeiteration2}, \texttt{LabelFrontier}, is similar to \texttt{LabelSeed}, but instead of checking if $e$ is the longest-edge in both adjacent triangles $t_a$ and $t_b$, this kernel checks if $e$ is not the longest-edge of any of the two  triangles. In that case $e$ is labeled as  frontier-edge and  used in the next phase to delimit a new polygon. In case $e$ is a boundary edge,  $e$ is labeled as a frontier-edge, independent of the size of $e$.

After the algorithm calls each kernel consecutively, 
%the algorithm has a label triangulation $\tau_L = (G,V)$ in global memory, now
each terminal-edge region $R_i$ is delimited by frontier-edges and a seed triangle per $R_i$ was stored as is shown in Figure \ref{fig:labeledtriangluation}. The data of the  triangulation $\tau_L = (G,V)$ is in global memory and ready to be used in the Traversal phase.

\subsubsection{Traversal phase}

In this phase,  as many polygons are built  as triangle indices are marked  \texttt{true} in the \texttt{Seed array}. This phase consists in just one kernel, the \texttt{TraversalTriangulation} kernel, and it is shown in Algorithm \ref{algo:TravelPhase}.

\begin{algorithm}
    \caption{Travel Phase}\label{algo:TravelPhase}
    \begin{algorithmic}[1]
    \Require Seed array $S$, Triangulation $\tau_L(V,E)$
    \Ensure Mesh array $M$ , Position array $I$.
    \BKernel{\texttt{TraversalTriangulation}($\tau = (V,E)$)}
        \ForAll{triangle $t_i \in$ $S$ \textbf{in parallel}} \label{algoltravel:triangleiteration}
            \State $P_i \leftarrow$ \texttt{PolygonConstruction}($t$, $\tau_L(V,E)$)
            \State \texttt{Synchronize threads}
            \State $i_{mesh} \leftarrow$ \texttt{AtomicAdd}( $i_{Gmesh}$ , $|P| + 1$)
            \State $i_{pos} \leftarrow$ \texttt{AtomicAdd}( $i_{Gpos}$ , $1$)
            \State $I[i_{pos}] \leftarrow i_{mesh}$  
            \State $M[i_{mesh}] \leftarrow |P|$  
            \For{$j \leftarrow 0$ \textbf{to} $|P|$}
                \State $M[i_{mesh} + 1 + j ] \leftarrow P[j]$
            \EndFor
        \EndFor
    \EKernel
    \end{algorithmic}
\end{algorithm}

\begin{figure}
    \centering
  \includegraphics[width=0.6\linewidth]{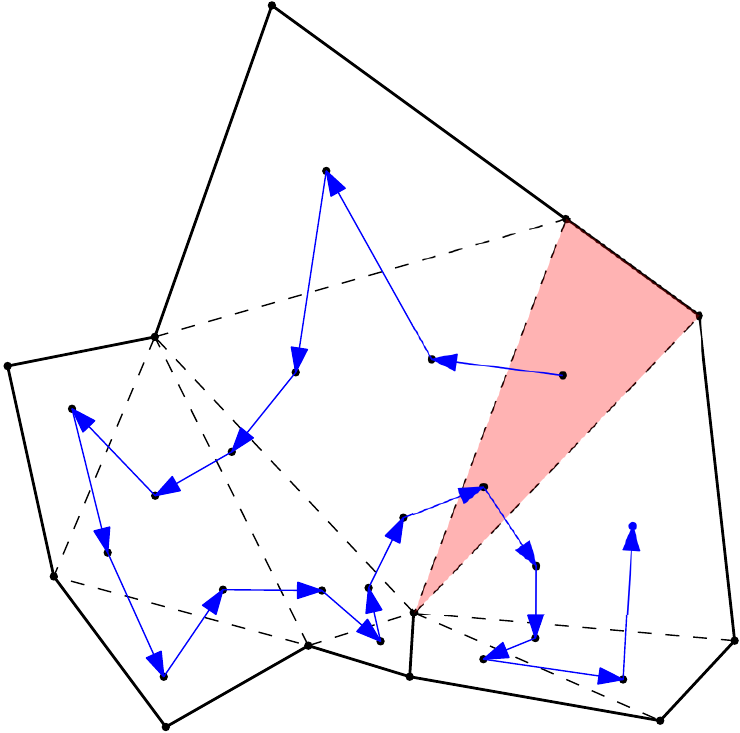}
  \caption{Example of a traversal inside of a terminal-edge region $R_i$ to the generation of polygon made by Algoritm \ref{algo:TravelPolyconstruction}. Red triangle is the seed triangle of $R_i$ and blue arrows are the traversal of the algorithm in $R_i$, for each triangle visited with frontier-edges, new edges are added as part of a new polygon.}
  \label{fig:traversal}
\end{figure}

The  \texttt{TraversalTriangulation} kernel uses the labeled triangulation $\tau_L$ of the previous phase to build and and store the polygon information in the \texttt{Mesh array} and the \texttt{Position array}. The kernel   assigns to each thread a triangle  in the \texttt{Seed array}. If $Seed[i]$  is \texttt{True},   triangle $t_i$ is used as starting triangle to build a polygon in \texttt{PolyConstruction} (see Algorithm \ref{algo:TravelPolyconstruction}). \texttt{PolyConstruction} travels inside a terminal-edge region $R_i$ in counter-clock-wise and stores the frontier-edges of each $t$ inside $R_i$ as part of the polygon $P_i$. As each vertex of $\tau$ is an endpoint of a frontier-edge, \texttt{PolyConstruction} visits all vertices and edges  in $R_i$. Note that if $t_i$ has 3 frontier-edge, $t_i$ is stored as the  polygon  $P_i$. An example of the traversal made by \texttt{PolyConstruction}  is shown in Figure \ref{fig:traversal}.

\begin{algorithm}
    \caption{Polygon construction}\label{algo:TravelPolyconstruction}
    \begin{algorithmic}[1]
    \Require Seed triangle $t$, Triangulation $\tau_L = (V,E)$
    \Ensure Polygon $P$ as a set of vertices.
    \Function{PolyConstruction}{$t$, $\tau_L=(V,E)$}
    \State $P$ $\leftarrow \emptyset$
    \State $v_{next}$, $v_{init}$ $\leftarrow$ Any vertex of $t$
    \State $t_{next}$, $t_{init}$ $\leftarrow$ t
    \Do
        \If{$t_{next}$ has frontier-edges}
            \State Store frontier-edge in $P$ in ccw
            \State $v_{next}$ $\leftarrow$ Last vertex store in $P$
            \State $t_{next}$ $\leftarrow$ Adjacent triangle by an internal-edge in ccw that share $v_{next}$.
        \EndIf
    \doWhile{$t_{next} \not= t_{init}$ \textbf{or} $v_{next} \not= v_{init}$}
    \State \Return P
    \EndFunction
    \end{algorithmic}
\end{algorithm}

After the polygon $P_i$ was generated and stored in a register,  the  \texttt{TraversalTriangulation} kernel stores $P_i$ in the \texttt{Mesh array}. Since several threads attempt to store concurrently their generated polygon in the \texttt{Mesh array}, the kernel manages a global index $i_{Gmesh}$ to get a unique index in \texttt{Mesh array}, each thread has to use to append a new polygon. The global index is obtained using  \textsc{AtomicAdd}. 
%first at  $i_{Gmesh}$ the length of a new  polygon plus one space is used to store the length of $|P_i|$ and their vertices. 

\textsc{AtomicAdd} returns the value of $i_{Gmesh}$ before the addition, therefore the kernel uses  this value as the index position in $i_{Mesh}$ where to store the number of  vertices and then the  polygon vertices. The $i_{mesh}$ value is stored in the \texttt{Position array} in the same way using the global index $i_{Gpos}$.

After the call to the \texttt{TraversalTriangulation} kernel, the algorithm already has a polygonal mesh $\tau' = (V,E)$, stored in the \texttt{Mesh array} and the \texttt{Position array}, but some of their polygons might be  non-simple. The next phase, the Reparation phase,  is in charge of partitioning non-simple polygons into simple ones. 

\subsubsection{Reparation phase}
\label{subsubsec:reparation}

In this phase the algorithm repairs non-simple polygons. To do so, the algorithm handles two auxiliary arrays to represent a temporal polygon mesh $\tau'_{aux} = (V,E)$: the \texttt{Aux Mesh array} and \texttt{Aux Position array}. The  \texttt{PolygonReparation} kernel shown in Algorithm \ref{algo:ReparationPhase}  splits  each non-simple polygon $P_i$ in two new polygons using a diagonal containing a  barrier-edge tip of $P_i$ and store the result in $\tau'_{aux}$. If $\tau'_{aux}$ still contains barrier-edge tips, \texttt{PolygonReparation} kernel with $\tau'_{aux}$ as input mesh  is called again. This process continues until all the barrier-edge tips are removed.

\begin{algorithm}
    \caption{Reparation Phase}\label{algo:ReparationPhase}
    \begin{algorithmic}[1]
    \Require Mesh array  $M_{in}$, Position array $I_{in}$, Labeled triangulation $\tau_L(V,E)$
    \Ensure Mesh array  $M_{out}$, Position array $I_{out}$
    %\Ensure Polygonal Mesh $\{ M_{out}, I_{out}\}$.
    \BKernel{ \texttt{PolygonReparation}($\{ M_{in}, I_{in} \}$, $\tau_L$) }
        \ForAll{polygon $P_i$ in $\{ M_{in}, I_{in} \}$  \textbf{in parallel}} \label{algorepa:polygoniteration}
            %\State Copy $p_i$ to register $P$
            \ForAll{consecutive vertices $v_a$, $v_b$, $v_c$ $\in P_i$}
                \If{$v_a$ equals $v_c$}
                    \State $v_{bet} \leftarrow v_b $
                    \State break
                \EndIf
            \EndFor
            \If{$v_{bet} \not= \emptyset$}
                \State $e_i \leftarrow$ Middle internal-edge incident to $v_{bet}$
                \State Label $e_i$ as frontier-edge
                \State Be $t_a$ and $t_b$ triangles that share $e_i$
                \State $P_a$ $\leftarrow$ \texttt{PolyConstruction}($t_a$, $\tau_L(V,E)$)
                \State $P_b$ $\leftarrow$ \texttt{PolyConstruction}($t_b$, $\tau_L(V,E)$)
                \State Store $P_a$ and $P_b$ in $M_{out}$
            \Else
                \State Store $P_i$ in $M_{out}$
            \EndIf
        \EndFor
    \State \texttt{AtomicAdd}(\texttt{Total bet}, $|v_{bet}|$)
    \EKernel
    \end{algorithmic}
\end{algorithm}

To repair each non-simple polygon in  $\tau'$, the kernel \texttt{PolygonReparation} assigns to each thread a polygon  $P_i$. Note that there is two ways to store the temporal polygons in this phase, one is using registers and the second one is reading and constructing the polygon directly in the auxilary mesh. Both ways are possible due to the length of the two new polygons is $|P_a| + |P_b| = |P| + 2$ in case of barrier-edge tip. % in known at moment that a barrier-edge tip% is detected as is proof in lemma \ref{lemma:sizepolygon}.

%\begin{lemma}\label{lemma:sizepolygon}
%For each polygon $P$ of size $|P|$ with barrier-edge tip. The reparation phase will generate two new polygon $P_a, P_b$ of size $|P_a| + |P_b| = |P| + 2$
%\end{lemma}

%\begin{proof}

%\textbf{Proof.} By construction, the reparation phase splits a polygon by adding a new edge $e = \{v_1, v_2\}$ between a barrier-edge tip of $P$ and an endpoint of an edge of $P$ generating two new polygons $P_a, P_b$ without overlap and with both containing $e$. Thus, $P_a \cup P_b = P$ and $P_a \cap P_b = \{v_1, v_2\}$ then:

%\begin{align*}
%|P_a \cup  P_b|  &= |P_a| + |P_b| - |P_a \cap P_b|     \\
%|P_a| + |P_b| &= |P_a  \cup  P_b| + |P_a \cap P_b| \\
%|P_a| + |P_b| &= |P| + 2
%\end{align*}

%Therefore the memory of the two new polygons generated in the reparation after the split is $|P| + 2$. $\Box$
%\end{proof}

The kernel first checks if $P_i$ contains a barrier-edge tip by comparing 3 consecutive vertices  $v_a$, $v_b$, $v_c \in P_i$. If $v_a = v_c$, then $v_b$ is a barrier-edge tip,  stored in $v_{bet}$, and used to start the process of repairing $P_i$. If the kernel does not  detect barrier-edge tips then $P$ is stored as part of the polygonal mesh $\tau'_{aux}$. .

%The process of repair refers to split the polygon $P_i$ in two polygons $P_a, P_b$ and store both in $\tau'_{aux}$. 

\begin{figure}
    \centering
  \includegraphics[width=0.6\linewidth]{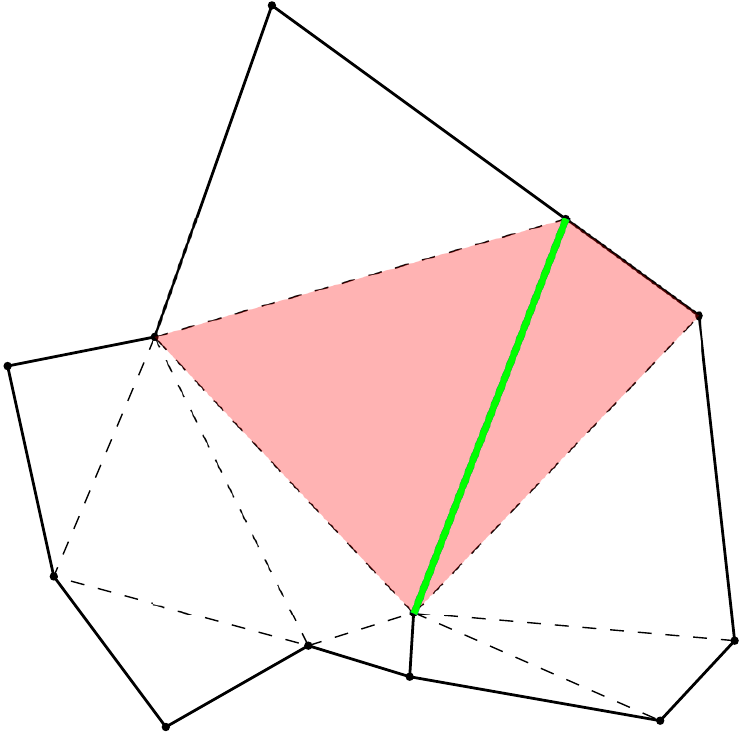}
  \caption{Example of reparation of polygon shown in Figure \ref{fig:traversal}. Green edge is the middle-edge choose to split the polygon. This edge is label as frontier-edge and their two adjacent triangles, in red, are uses to generate two new polygons by calling to Algorithm \ref{algo:TravelPolyconstruction}. }
  \label{fig:polysplot}
\end{figure}

Starting from a barrier-edge tip $v_{bet} \in P_i$,  the algorithm searches for an incident triangle to $v_{bet}$ using the \texttt{Trivertex} array.  Through this triangle, all internal edges (diagonals) with $v_{bet}$ as endpoint are looked for and the middle internal-edge  $e$ is  labeled  as a new frontier-edge. This strategy divides $P_i$ into two new polygons $P_a$ and $ P_b$. The two adjacent triangles  $t_a, t_b$ to $e$ are stored and used as seed triangles to generate $P_a, P_b$ by calling Algorithm \ref{algo:TravelPolyconstruction}. This process is shown in Figure \ref{fig:polysplot}. To store the polygons in $\tau'_{aux}$,  \textsc{AtomicAdd} is used again with two global indices in the same way as in Algorithm \ref{algo:TravelPhase}. 
%Note that the use of \texttt{Trivertex} is optional since we can search in the \texttt{triangle array} an triangle incident to $v_{bet}$ but ti would cost $O(m)$ time per barrier-edge tip, with $m$ the number of triangles in $\tau = (V,E)$. 

At the end of the kernel,  \textsc{AtomicAdd}is called again  with the global variable \texttt{Total bet}, to check if there are further barrier-edge tips. The kernel \texttt{Total bet} \texttt{PolygonReparation} is called again until \texttt{Total bet} gets the value $0$.

The final output is polygonal mesh $\tau' = (V,E)$ with simple polygons of arbitrary shape.

\section{Experiments}
\label{sec:Experiments}

\begin{figure}
  \includegraphics[width=1\linewidth]{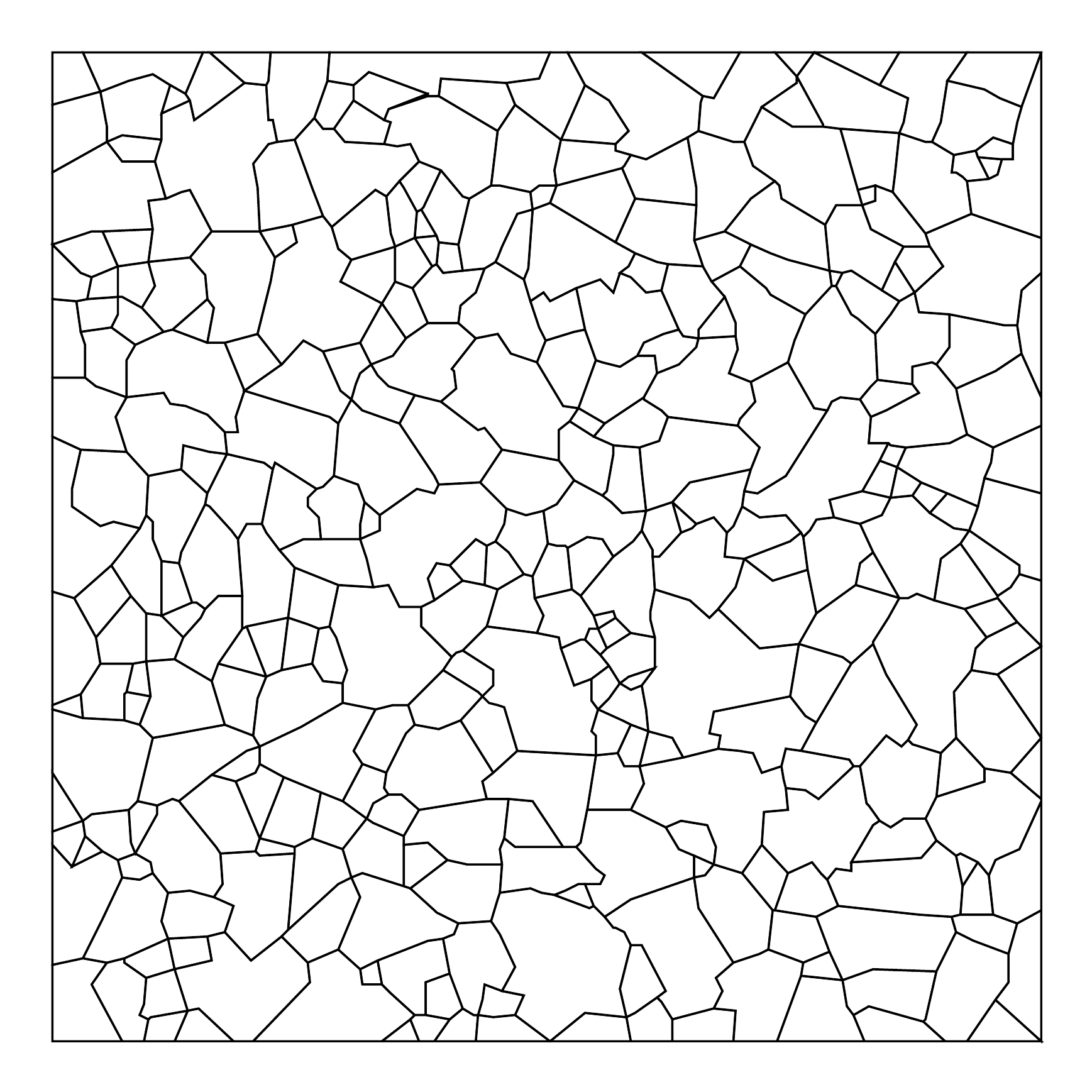}
  \caption{Kind of polygonal mesh generated in the experiments. This mesh was generated using 10000 random poins.}
  \label{fig:examplemesh}
\end{figure}

\begin{figure}
  \includegraphics[width=1.15\linewidth]{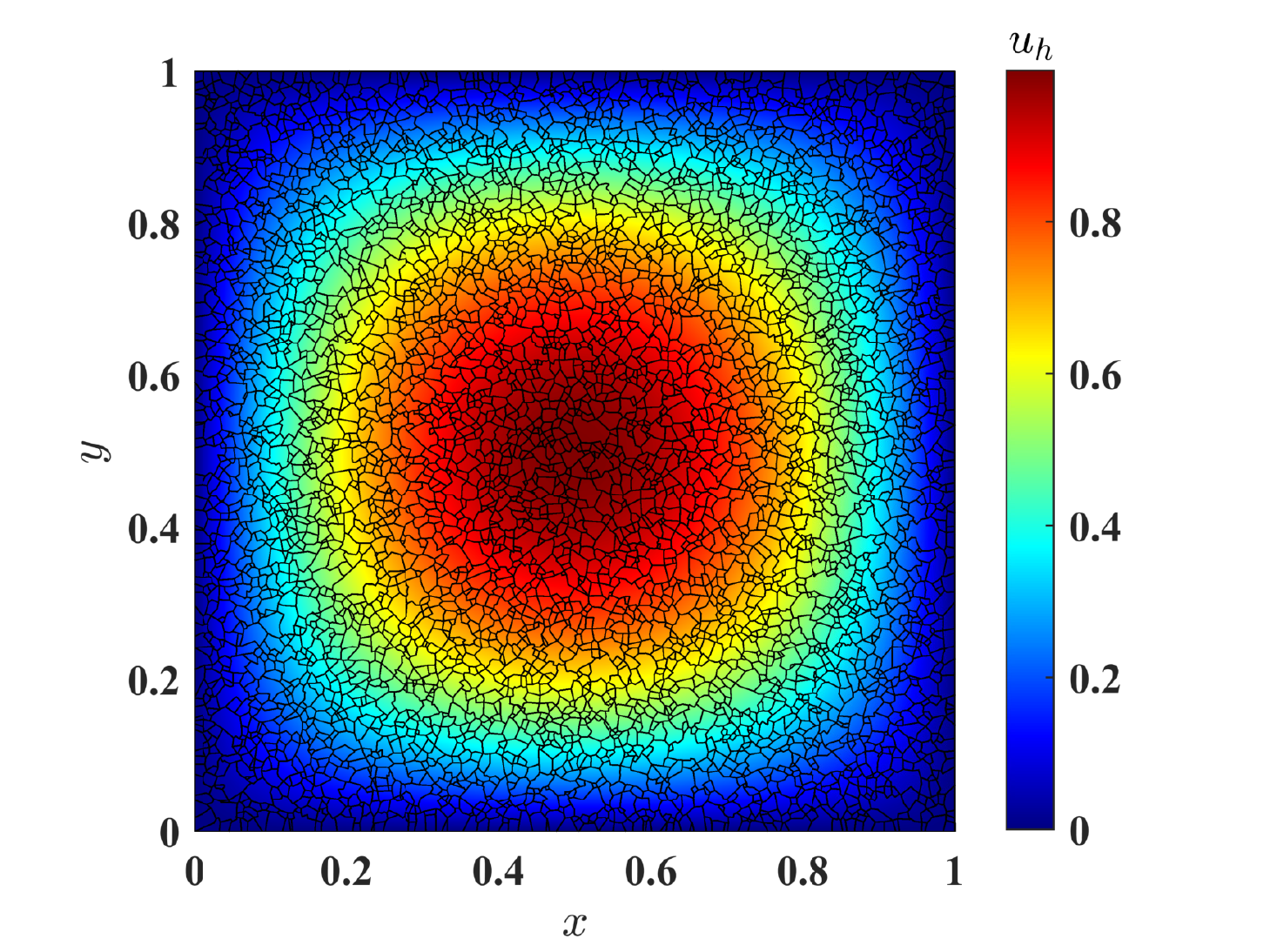}
  \caption{Preliminary simulation results using a polygonal mesh and  the Virtual Element Method for  a problem modeled with the Laplace equation.}
  \label{fig:Disk}
\end{figure}

For testing the algorithm, we used a computer with a CPU Intel(R) Core(TM) i5-9600K of 3.70GHz and a GPU NVIDIA GeForce RTX 2060 SUPER. The algorithm was programmed  in CUDA C++.  The  input are random point sets inside a square of sides 10000x10000 stored in point files {\sc .node}. We used the software Triangle \cite{triangle2d} to generate a Delaunay triangulation with the parameters \textsc{-zn} in  the triangle file {\sc .ele} and the neighbor file {\sc .neigh}. In addition, we created a script to generate the triangle per vertex file {\sc .trivertex} to accelerate the reparation phase. All polygonal meshes were generated using $65535$ blocks and $1024$ threads per block.  An example of  the kind of  meshes generated in  the experiments is shown in Figure \ref{fig:examplemesh}. We also did a preliminary assess of the resulting polygonal meshes with the virtual element method (VEM) and the  result is shown in Figure \ref{fig:Disk}.

\begin{figure}
  \includegraphics[width=\linewidth]{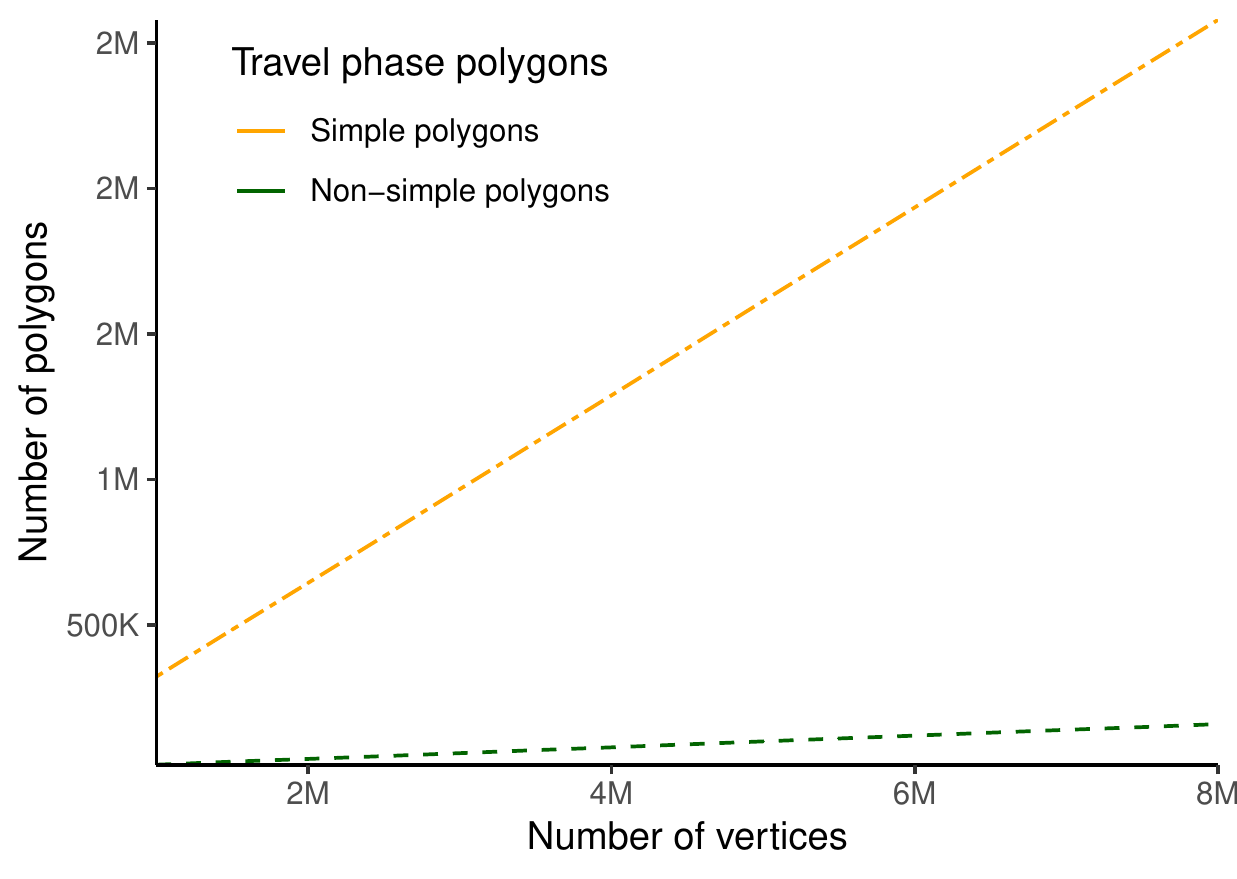}
  \caption{Kind of polygons generated after the travel phase shown in Algorithm \ref{algo:TravelPhase}}
  \label{fig:travelphasepoly}
\end{figure}

\begin{figure}
  \includegraphics[width=\linewidth]{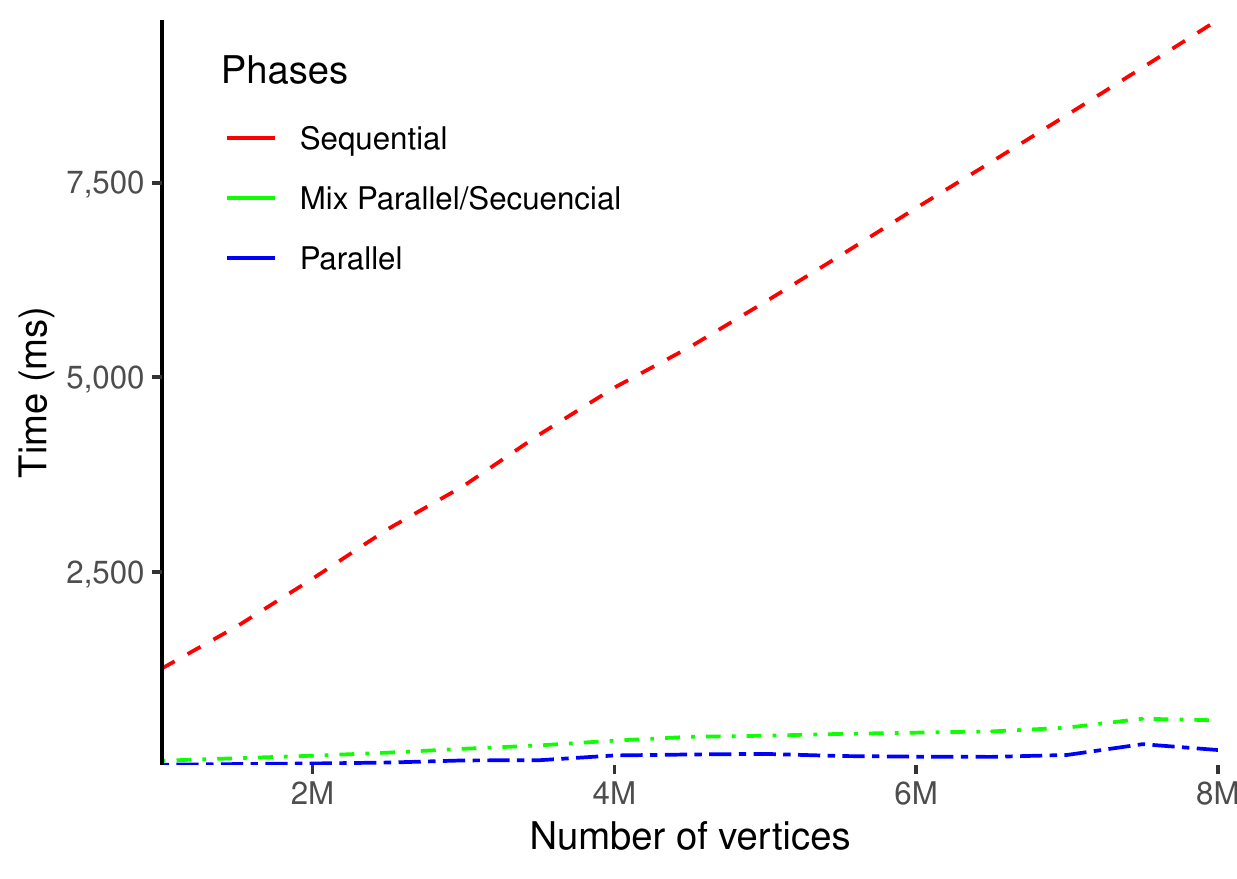}
  \caption{Comparison between the time of the sequential version of the algorithm and the parallel version}
  \label{fig:timecomp2}
\end{figure}

We used  the same input trianngulations, varying from 1 million to 8 million points, to compare  the time performance of the parallel version, the sequential version and a mixed version. The mixed  version uses the parallel version of the label phase and Traversal  phase and the sequential version of the Reparation phase. Each test was repeated five times; the average time of the experiments is shown in Figure \ref{fig:timecomp2}. As we said previously, traversal phase can generate simple and non-simple polygons. The number of  non-simple polygons generated in the experiments and needed to be repaired are shown in Figure \ref{fig:travelphasepoly}.

\begin{figure}
  \includegraphics[width=\linewidth]{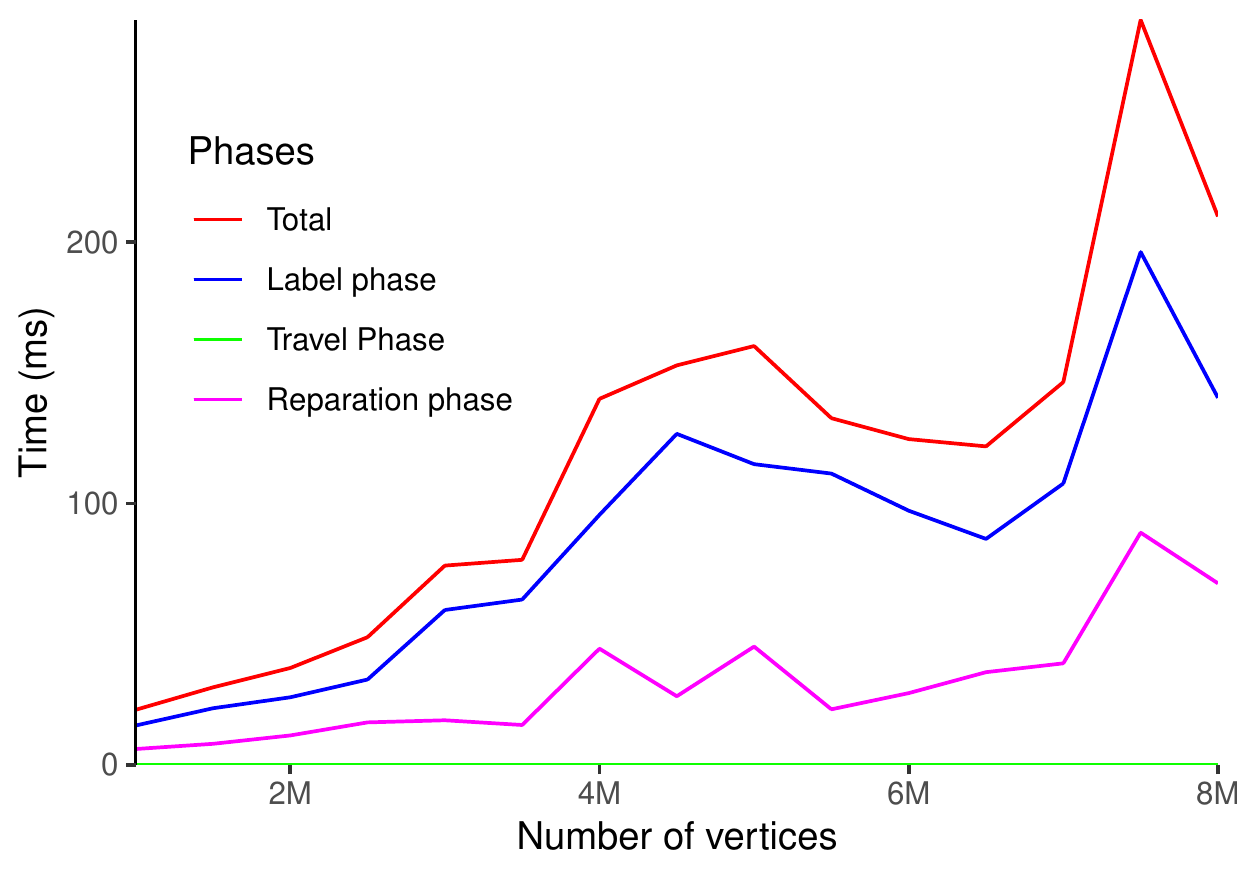}
  \caption{Comparison between the time of each phase in parallel, ``Total" refers to the sum of the time of all phases.}
  \label{fig:phasecomp}
\end{figure}

\begin{figure}
  \includegraphics[width=\linewidth]{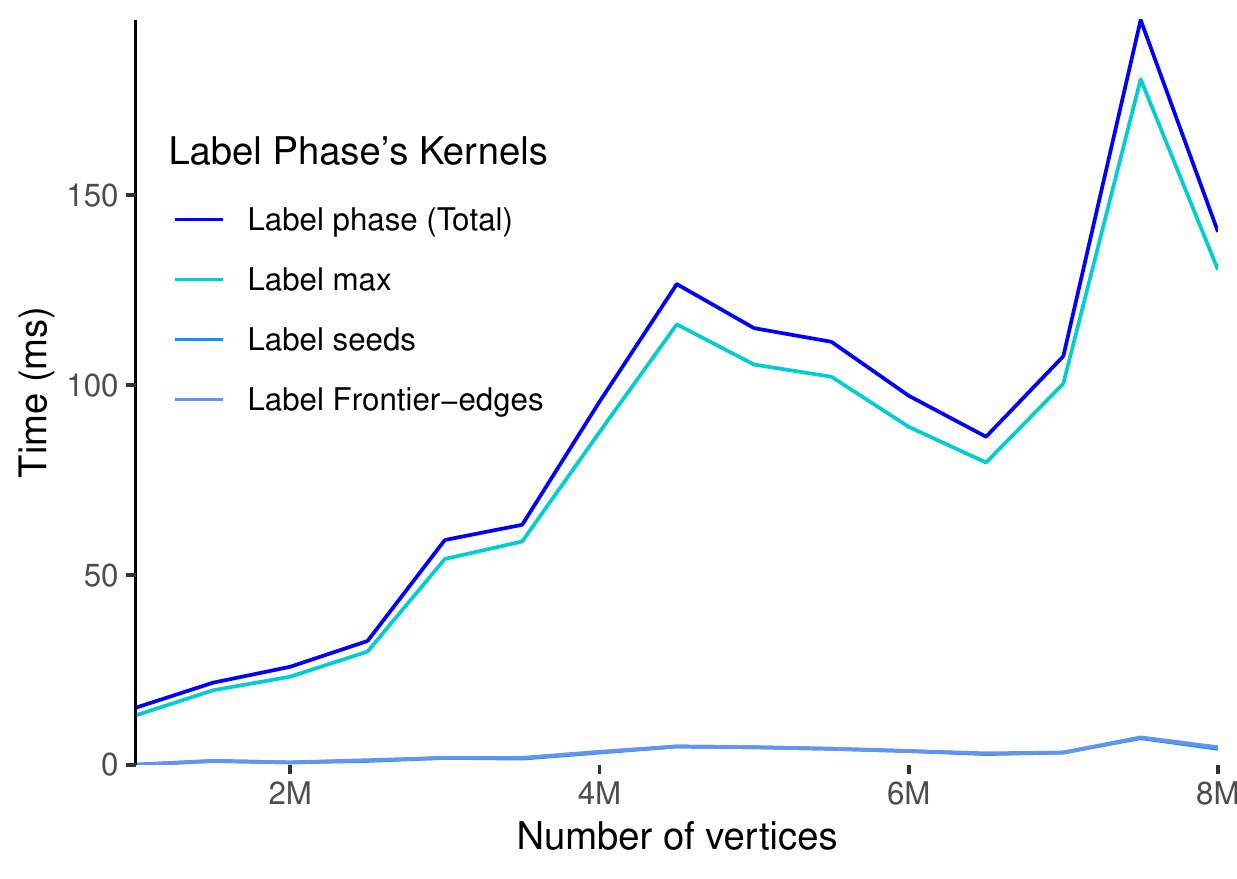}
  \caption{Comparison between the $3$ kernels that are called in the label phase and where shown in Algorithm \ref{algo:labelphase}.}
  \label{fig:labelphasecomp}
\end{figure}

The average time of each phase is shown in Figure \ref{fig:phasecomp}. The faster phase is the traversal phase; this is the simplest kernel since it just calls  one kernel once and  is applied to a subset of triangles  instead to the whole domain. The phase that takes longer was the Label phase. This phase consist in the call to $3$ kernels applied to the edges and triangles of whole triangulation. Figure \ref{fig:labelphasecomp} shows the time of each kernel; the slowest kernel is the \texttt{Labelmax} kernel. This can be explained due to the floating-point arithmetic used to calculate the length of each edge of the input triangulation. The most inefficient parallel phase is the reparation phase, as shown in Figure \ref{fig:travelphasepoly}. The number of polygons to repair  are  very few in contrast with the time that this phase takes. There is no significant advantage of using  the parallel version   over the sequential version in this phase.

\section{Conclusions and future work}
\label{sec:conclusions}
We have presented an algorithm to build a polygonal mesh formed by convex and non-convex polygons in parallel on GPU-architectures. We hope that this kind of polygonal meshes can be used as an alternative of constrained Voronoi  meshes in polygonal finite element methods such as the VEM. Note that Voronoi cells must be cut against the boundary of the input geometry. Terminal-edge regions are always inside the domain. 

Our preliminary experimental results show that the GPU programming model allow us to accelerate the sequential version but we still want  to do further experiments before computing  the speedup. Memory improvements are also necessary in order to minimize the number of read and write operations over the global device memory. The possibility of using shared memory to manage the polygonal mesh should also be evaluated. Using {\sc AtomicAdd} to manage the index of the polygonal mesh between concurrent  threads works with meshes of size inferior to $8$ millions of vertices, but we still have  to solve memory problems to generate larger meshes.

We believe that some of the ideas implemented in the Label and Traversal phase can be adapted to accelerate algorithms based on the Longest-edge propagation path on GPU.

\bibliography{demoref}

\end{document}